\documentclass[sn-mathphys-num, pdflatex]{sn-jnl}%

\usepackage{graphicx}%
\usepackage{multirow}%
\usepackage{amsmath,amssymb,amsfonts}%
\usepackage{amsthm}%
\usepackage{mathrsfs}%
\usepackage[title]{appendix}%
\usepackage{xcolor}%
\usepackage{textcomp}%
\usepackage{manyfoot}%
\usepackage{booktabs}%
\usepackage{algorithm}%
\usepackage{algorithmicx}%
\usepackage{algpseudocode}%
\usepackage{listings}%
\usepackage[T1]{fontenc}%
\usepackage[utf8]{inputenc}%
\usepackage{babel}%
\usepackage{textcomp}%
\usepackage{upgreek}%
\usepackage{subcaption}
\begin{document}

\title{Development of a novel Light and Ion Beam Induced Luminescence (LIBIL) setup for in-situ optical characterization of color centers in diamond}

\author[1,2]{\fnm{Matija} \sur{Matijević}}
\equalcont{These authors contributed equally to this work.}

\author[1]{\fnm{Livio} \sur{Žužić}}
\equalcont{These authors contributed equally to this work.}

\author[2]{\fnm{Jacopo} \sur{Forneris}}

\author*[1]{\fnm{Zdravko} \sur{Siketić}}\email{zsiketic@irb.hr}

\affil[1]{\orgdiv{Laboratory for ion beam interactions}, \orgname{Ruđer Bošković Institute}, \orgaddress{\street{Bijenička cesta 54}, \postcode{10000}, \city{Zagreb}, \country{Croatia}}}

\affil[2]{\orgdiv{Department of Physics}, \orgname{University of Turin}, \orgaddress{\street{Via Pietro Giuria 1}, \postcode{10125}, \city{Turin}, \country{Italy}}}

\abstract{In this work, development of the new Laser and Ion Beam Induced Luminescence (LIBIL) experimental end-station has been presented. To systematically test the capabilities and limitations of the newly developed setup, ionoluminescence (IL) and iono-photoluminescence (IPL) measurements were performed on a type IIa optical grade and a type Ib nitrogen rich diamond. By comparing and analyzing the obtained spectra, it was shown that the speed of luminescence quenching has a non-trivial dependence on the ion beam current. Additionally, it was demonstrated that some spectral features characteristic of the negatively charged nitrogen-vacancy color center (i.e. NV$^-$ zero-phonon line) have been observed only during IPL measurements. This demonstrates that the unification of two separate steps, ion implantation and optical characterization,  could yield new insights into dynamics of color center formation.}

\maketitle

\section{Introduction}

Color centers in diamond are crystalline defects that are responsible for imparting color to an otherwise transparent crystal. Although studied for decades, in recent years interest in them has surged due to their potential applications in quantum sensing and quantum computing\cite{awschalom2018}. One such defect, the nitrogen-vacancy (NV) color center, exists in two optically active charge states: negatively charged (NV$^-$), and neutral (NV$^0$)\cite{doherty2013}. The NV$^-$ center is especially interesting due to it's reported photostability, and ability to optically initiate, manipulate, and measure its spin state with high fidelity at room temperature\cite{kurtsiefer2000, brouri2000, jelezko2004}. This makes it a promising candidate for quantum network qubits. Beyond quantum information processing, NV$^-$ centers offer numerous quantum metrology applications, such as: magnetometry\cite{taylor2008}, electrometry\cite{dolde2014}, barometry\cite{doherty2014}, and thermometry\cite{acosta2010}.

Synthetic diamonds, can be made using two different procedures: High Temperature and High Pressure (HPHT) and Chemical Vapor Deposition (CVD). In HPHT diamond growth, crystals are generated by dissolving a carbon material (i.e. graphite) with a flux of molten metal (usually iron, nickel or cobalt). The carbon is then deposited on a substrate containing diamond seed crystals in a specialized press generating pressures of around 6 GPa and temperatures of 1500 \textdegree C\cite{sonin2022} replicating the conditions of natural diamond growth. CVD on the other hand, is a method that uses a carrier gas (usually a hydrogen-methane mix) to transport constituent elements (carbon) onto a heated substrate containing nucleation centers (i.e. diamond seeds)\cite{may2000}. Both of these methods allow for creation of single crystal diamonds with controlled concentrations of impurities (down to a few ppb).

Ion implantation is a technique in which different ion species are used to irradiate solid targets which modifies their physical, chemical and electrical properties. It is a well known technique in the semiconductor industry for incorporation of dopants in a material\cite{williams1998}. Ion implantation offers many advantages for defect engineering over incorporation of defects during growth. Ion beams can be focused down to nanometer sized spots\cite{rubanov2011}, allowing for laterally resolved color center creation. By tuning the ion energy it is possible to precisely define the penetration range of ions in a material, allowing for precise longitudinal placement of color centers. Furthermore, by selecting the implanted ion species it is possible to create color centers with varying optical properties\cite{bradac2019}. Finally, precise ion counting systems allows for well defined determination of defect concentration\cite{vicentijevic2023}. The ion microprobe setup, having all the capabilities mentioned above, is an ideal choice for defect engineering by ion implantation\cite{nietohernandez2024}.

Ionoluminescence (IL) or Ion Beam Induced Luminescence (IBIL) is an ion beam analysis (IBA) technique employed for material investigation\cite{bettiol1994, manfredotti2001, calvodelcastillo2007}. It can be paired with other IBA techniques such as Particle Induced X-ray Emission (PIXE) for trace element analysis\cite{logiudice2003}, or Ion Beam Induced Charge (IBIC) for analyzing radiative recombination centers\cite{manfredotti1998}. Furthermore, it can be used to excite optically active defects\cite{markovic2015}, similarly to off-resonant laser excitation in confocal fluorescence microscopy, which is often used for characterization of such defects (e.g. 532 nm laser for NV$^0$ \& NV$^-$ centers)\cite{schirhagl2014}. However, to obtain a detectable signal from a sample using a typical IL setup, a high ion current is required resulting in rapid accumulation of damage. This hinders tracking of defects like NV centers and leading to the formation of more complex, potentially non-luminescent defect structures. To address the above mentioned limitation, laser excitation is paired with ion irradiation which enables in-situ tracking of color center formation processes in diamond (and other materials). This provides crucial insight into the evolution of color centers which is paramount for realizing the full potential of color center based devices. From now on, due to the combined nature of excitation we will call this new technique Laser and Ion Beam Induced Luminescence (LIBIL), or alternatively, iono-photoluminscence (IPL).

In this work we will present the development and performance of a LIBIL end-station developed at Ruđer Bošković Institute (RBI) in Zagreb, Croatia.

\section{Methods}

The Ruđer Bošković Institute accelerator facility consists of two electrostatic accelerators: a 1 MV Tandetron, and a 6 MV EN tandem Van de Graaff. Each accelerator is coupled to two ion sources: a plasma source with charge exchange for extraction of He$^-$ ions, and a cesium sputtering cathode source for extraction of almost all other ions (from H$^-$ to Au$^-$). This allows the facility to provide a wide range of ion species and energies into a total of nine different experimental end-stations. Five beam lines can accommodate ions beams from either accelerator, two are able to work with ion beams from both accelerators simultaneously, and the rest are coupled only to the 1 MV Tandetron accelerator and its corresponding ion sources. Two of the nine mentioned end-stations have microbeam focusing capabilities.

\begin{figure}
	\includegraphics[width=\textwidth]{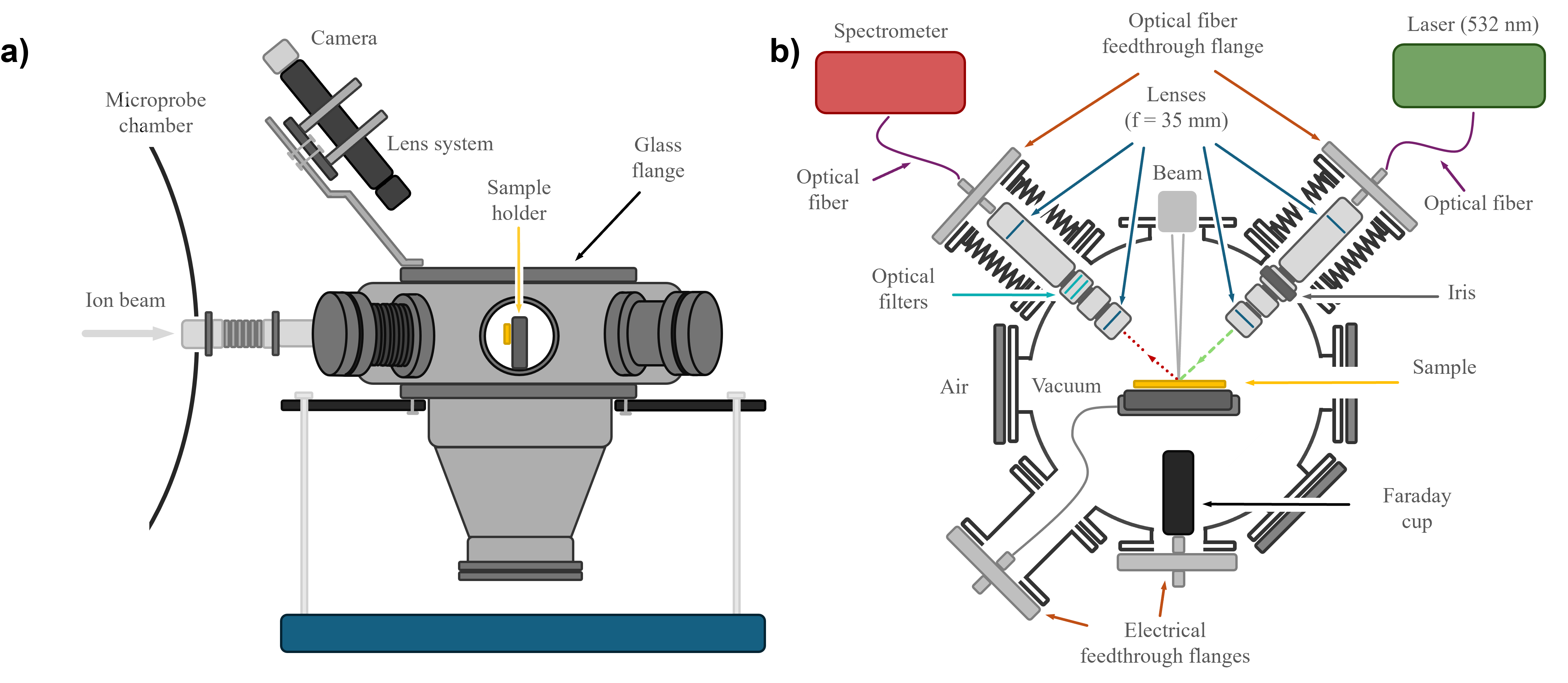}
	\caption{Schematic of the experimental end-station: \textbf{a)} Side-view \textbf{b)} Top-view}
	\label{fig1}
	\centering
\end{figure}

The LIBIL end-station is placed in extension of the ion microbeam line\cite{jaksic2007}. It is made up of a cylindrical vacuum chamber which hosts two symmetrically placed optical systems, one coupled to a green continuous wave laser (Integrated Optics CW 532 nm SLM Laser), and the other to a spectrometer (Ocean Optics HR4000). Each optical system consists of two lenses (Uncoated UV Fused Silica Plano-Convex, $\varnothing=25.4\ \mathrm{mm}$, $f=35\ \mathrm{mm}$), for focusing/collecting light, and a fiber feedthrough flange ($\varnothing=200\ \mathrm{\upmu m}$), which is schematically shown in Fig. \ref{fig1}. Additionally, the laser optical system has an iris in order to reduce the beam halo and laser power on the sample. The spectrometer system has two spectral filters (Hard-Coated 550 nm Longpass and 700 nm Shortpass) in order to filter out wavelengths outside the desired spectral range. The samples are mounted on a nanometric precision positioning system (3 DoF SmarAct piezo stage), and are observed by an imaging system consisting of a Navitar 12x Zoom lens tube and Basler ace camera with a Field of View (FOV) of $1420\times990\ \mathrm{\upmu m^2}$. The arriving beam current is measured in the experimental chamber using a Faraday cup placed on a flange opposite the beam entrance, and the deposited fluence is calculated with an experimental error of around 10 \%.

To systematically test the luminescence detection capabilities of this newly developed setup, two different nitrogen-rich, single crystal diamonds were used in order to directly create a detectable amount of optically active defects even without a post-irradiation annealing treatment\cite{lake2021}. The sample used were:
\begin{itemize}
	\item $5\times5\times0.15\ \mathrm{mm^3}$ CVD synthesized, type IIa, optical grade, single crystal diamond grown in $\left<100\right>$ orientation with nitrogen concentration $<1\ \mathrm{ppm}$;
	\item $3\times3\times0.3\ \mathrm{mm^3}$ HPHT synthesized, type Ib, single crystal diamond grown in $\left<100\right>$ orientation with nitrogen concentration < 200 ppm (made by Element Six).
\end{itemize}
Both samples were irradiated with a 2 MeV H$^+$ ion beam. This specific ion and energy was chosen due to it's damage profile per ion being much lower than for other species with similar energies. For comparison, the optical signal was detected using both IL and IPL techniques. Focusing of the ion beam was done using a standard triplet magnetic quadrupole configuration to a spot size of around $10\times10\ \mathrm{\upmu m^2}$ and then raster scanned over an area of $250\times250\ \mathrm{\upmu m^2}$. Furthermore, the beam scan speed and the integration time of the spectrometer were coupled such that one spectrum is acquired per beam pass over the scan area. Luminescence measurements on both samples were carried out for different ion currents ranging from 25 to 600 pA, and laser powers between 1 and 32 mW.

Since the setup allows detection of a luminescence signal during irradiation, it is possible to extract some useful insight on dynamics of the luminescence signal coming from different optically active defects. The Birks-Black model\cite{birks1951}:
\begin{equation}
	I(F)=\frac{I_0}{1+F/F_{1/2}},
	\label{eq1}
\end{equation}
has first been developed to describe signal quenching due to ion irradiation in organic scintillators. Since then, it has been shown to be a good approximation of luminescence decay in optically active defects\cite{manfredotti2010, markovic2015}. By fitting the data to Eq. \ref{eq1} it is possible to evaluate a parameter $F_{1/2}$, which represents a fluence at which the intensity of the luminescence signal drops to one half it's pre-irradiation value and is dependent on: defect structure, and ion beam species and energy. In this work we have evaluated this parameter for several different irradiation runs performed on the aforementioned diamonds.

\section{Results and Discussion}

\begin{figure}[H]
	\includegraphics[width=\textwidth]{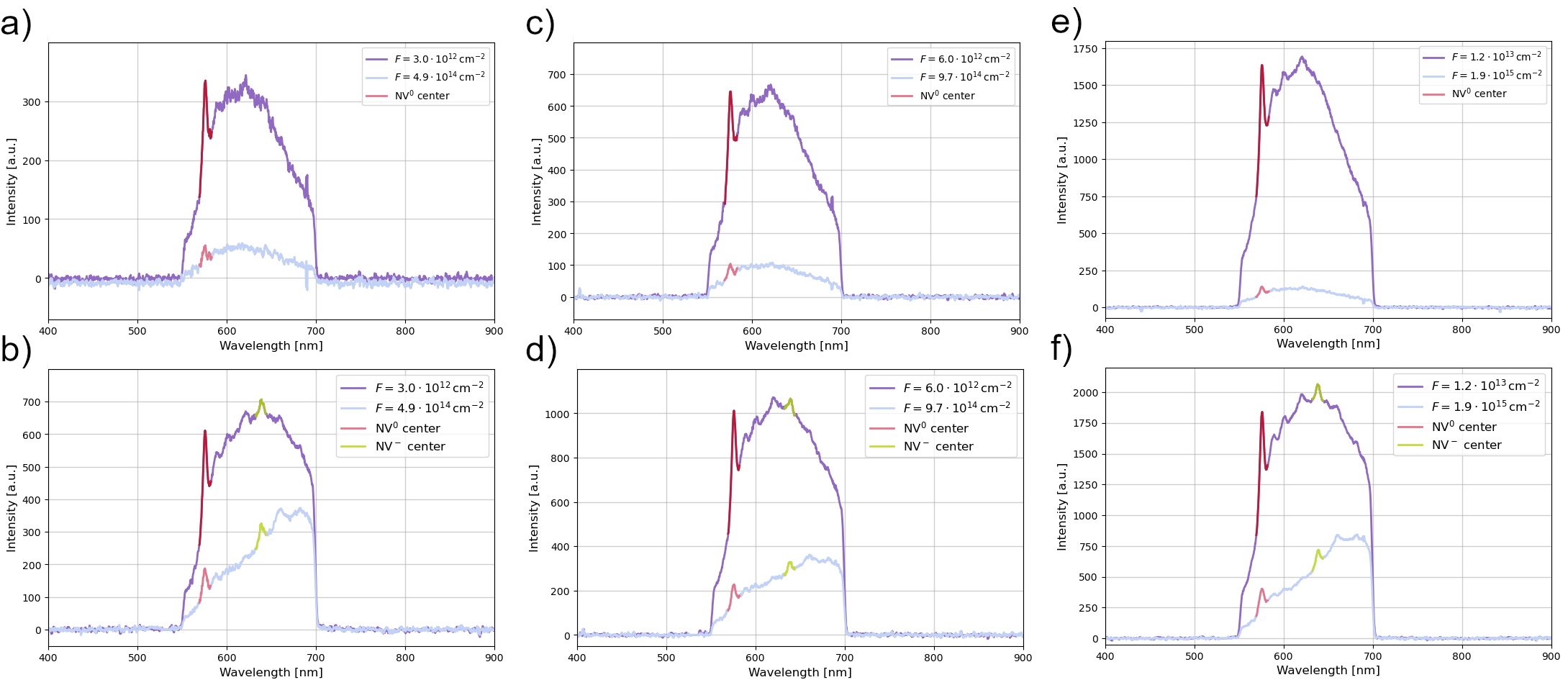}
	\caption{Spectra collected during 6 different IL and IPL runs on a type IIa optical grade CVD diamond: \textbf{a)} IL spectra, $I_\mathrm{ion}=50\,\mathrm{pA}$; \textbf{b)} IPL spectra, $I_\mathrm{ion}=50\,\mathrm{pA}$, $P_\mathrm{laser}=4.2\,\mathrm{mW}$; \textbf{c)} IL spectra, $I_\mathrm{ion}=100\,\mathrm{pA}$; \textbf{d)} IPL spectra, $I_\mathrm{ion}=100\,\mathrm{pA}$, $P_\mathrm{laser}=4.2\,\mathrm{mW}$; \textbf{e)} IL spectra, $I_\mathrm{ion}=200\,\mathrm{pA}$; \textbf{f)} IPL spectra, $I_\mathrm{ion}=200\,\mathrm{pA}$, $P_\mathrm{laser}=4.2\,\mathrm{mW}$.}
	\label{fig2}
	\centering
\end{figure}

In order to see if there is any increase in the collected luminescence signal, both IL and IPL measurements were performed on the type IIa optical grade diamond for ion beam currents of: 50, 100, and 200 pA. Fig. \ref{fig2} shows first and last spectra acquired by the setup during irradiation both with and without laser excitation. Just by comparing the spectra collected for comparable ion currents, it is clear that there is a distinct NV$^-$ peak visible in the spectra where laser excitation is present. Since it has been established that the NV$^-$ center is not optically active as a trap/recombination center, as was reported in other IL\cite{jia2025}, electroluminescence (EL)\cite{forneris2017}, and cathodoluminescence (CL)\cite{chen2024} experiments, it shouldn't be possible to observe NV$^-$ emission unless some articulated and proactive charge stabilization is applied\cite{haruyama2023}. This, along with the results shown in Fig. \ref{fig2}, shows that the LIBIL setup could resolve some spectral features not distinguishable only with ion excitation. Also, by comparing the overall intensity of emission during both IL and IPL measurements for comparable ion currents there is an increase of in overall emission intensity for IPL vs. IL measurements.

\begin{figure}
	\includegraphics[width=\textwidth]{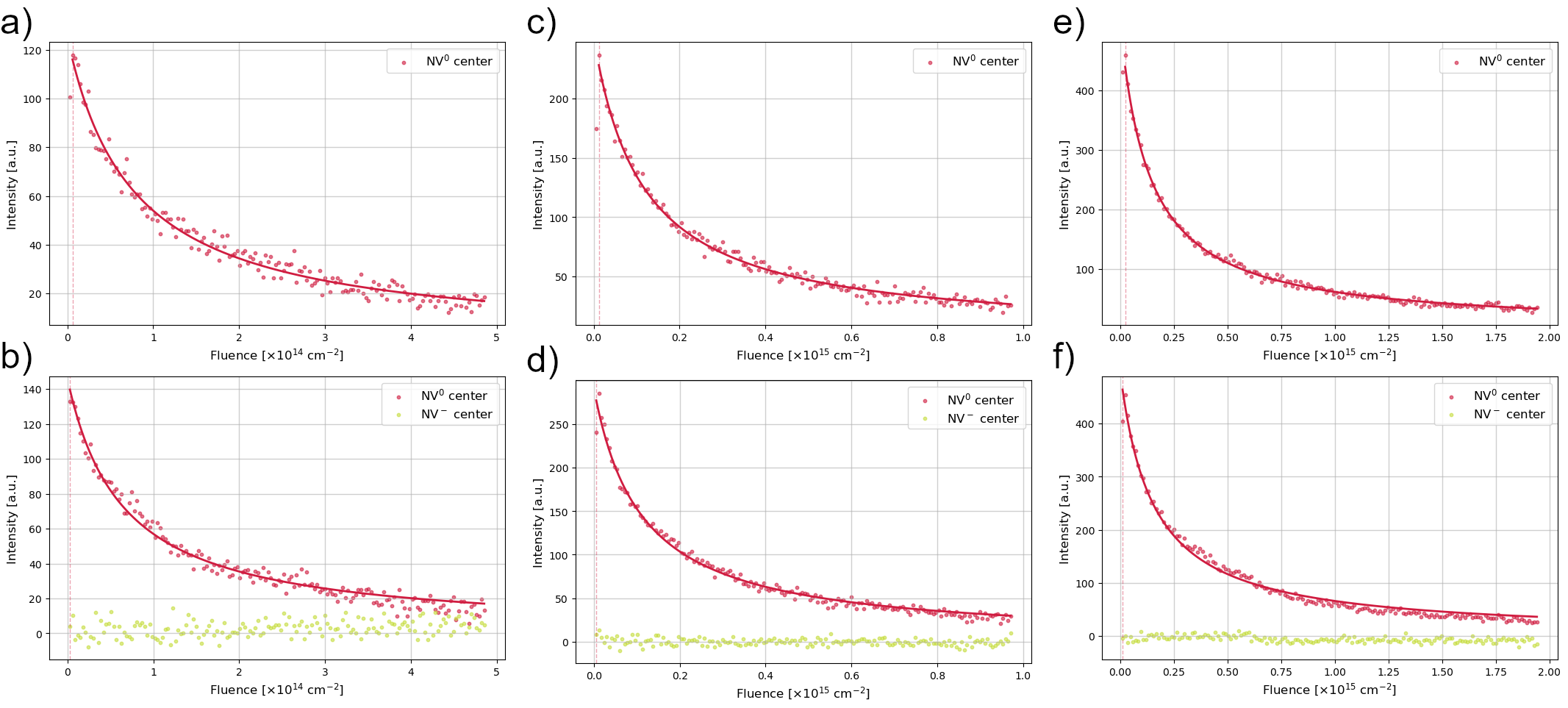}
	\caption{NV$^0$ and NV$^-$ intensity during 6 different IL and IPL runs on a type IIa optical grade CVD diamond: \textbf{a)} IL spectra, $I_\mathrm{ion}=50\,\mathrm{pA}$; \textbf{b)} IPL spectra, $I_\mathrm{ion}=50\,\mathrm{pA}$, $P_\mathrm{laser}=4.2\,\mathrm{mW}$; \textbf{c)} IL spectra, $I_\mathrm{ion}=100\,\mathrm{pA}$; \textbf{d)} IPL spectra, $I_\mathrm{ion}=100\,\mathrm{pA}$, $P_\mathrm{laser}=4.2\,\mathrm{mW}$; \textbf{e)} IL spectra, $I_\mathrm{ion}=200\,\mathrm{pA}$; \textbf{f)} IPL spectra, $I_\mathrm{ion}=200\,\mathrm{pA}$, $P_\mathrm{laser}=4.2\,\mathrm{mW}$.\newline
	The y-axis represents the difference in intensity of collected spectra from the PL spectrum acquired before irradiation.}
	\label{fig3}
	\centering
\end{figure}

\begin{figure}
	\centering
	\includegraphics[width=\textwidth]{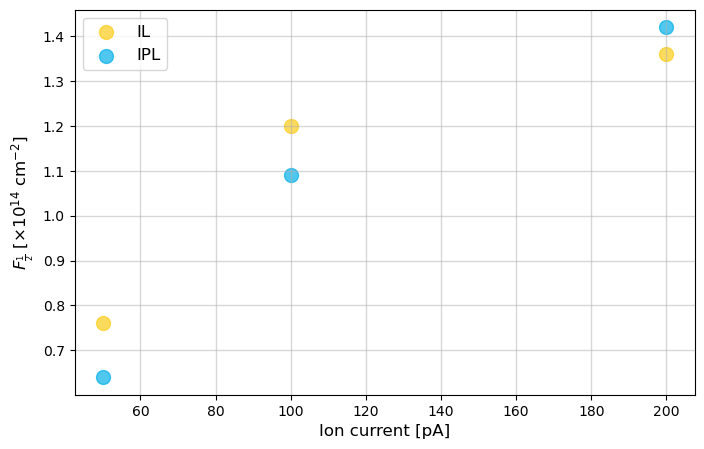}
	\caption{Graph displaying $F_{1/2}$ dependency on ion current.}
	\label{fig4}
\end{figure}

Intensity of the NV$^0$ zero-phonon line (ZPL) at 575 nm is calculated by integrating the observed peak. Results are plotted with respect to deposited fluence, as shown in Fig. \ref{fig3}. From the results it is clearly visible that intensity drops with increasing fluence, which is consistent with the previous statement that as damage accumulates, luminescence quenching defect complexes are formed. Data was fitted to Eq. \ref{eq1} to evaluate the values of the parameter $F_{1/2}$. The obtained results are graphed with respect to ion current in Fig. \ref{fig4} with IL results shown to be comparable to previously reported values found in literature\cite{markovic2015, manfredotti2010}. Furthermore, it appears that for lower ion currents the parameter decreases. This suggests that the luminescence quenching effect is more pronounced for lower ion beam currents which wasn't expected.  From this, we can conclude that the parameter $F_{1/2}$ has a non-trivial dependency on ion beam current. As no significant change in the NV$^-$ luminescence was observed, the parameter $F_{1/2}$ could not be evaluated in this case. The addition of laser excitation to IL measurements doesn't change the values of $F_{1/2}$ for comparable ion currents. This suggests that the introduction of laser excitation doesn't introduce further damage to the material. Taking everything into account, it has been shown that the LIBIL setup can help with detecting optically active defects that are below the detection limit for IL experiments.

\begin{figure}
	\includegraphics[width=\textwidth]{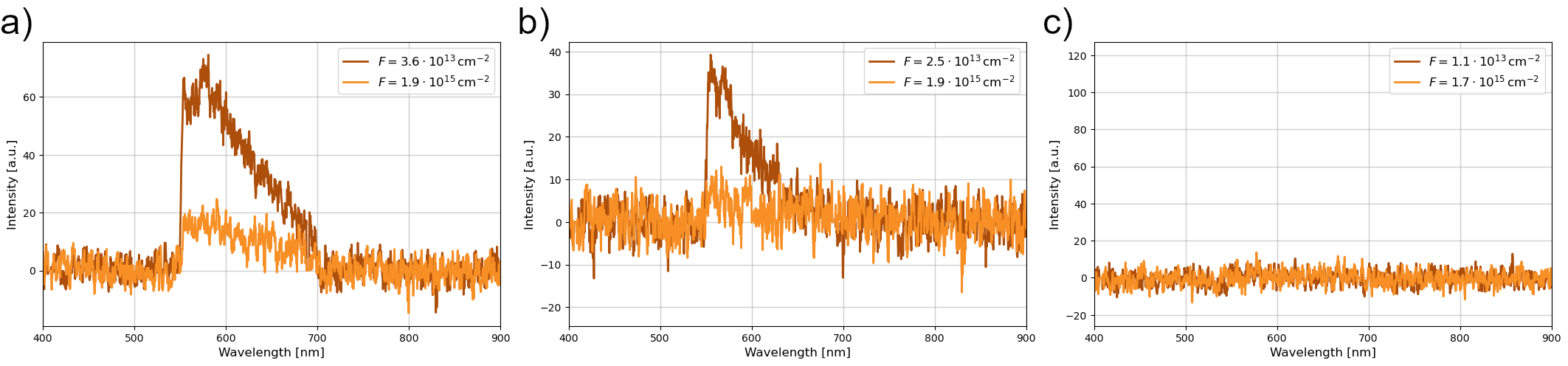}
	\caption{Spectra collected during 3 different IL runs on a type Ib HPHT diamond:\newline
	\textbf{a)} $I_\mathrm{ion}=600\,\mathrm{pA}$; \textbf{b)} $I_\mathrm{ion}=400\,\mathrm{pA}$; \textbf{c)} $I_\mathrm{ion}=200\,\mathrm{pA}$.}
	\label{fig5}
	\centering
\end{figure}

In the next step, IL measurements were performed on a type Ib diamond with ion beam currents of: 600, 400, and 200 pA. Fig. \ref{fig5} shows spectra collected by the setup during IL measurements. For 600 and 400 pA runs (Figs. 4a, b), there are no resolvable features associated with NV centers. Furthermore, it is clearly visible that no discernible IL signal was observed for the ion beam current of 200 pA (Fig 4c).

\begin{figure}
	\includegraphics[width=\textwidth]{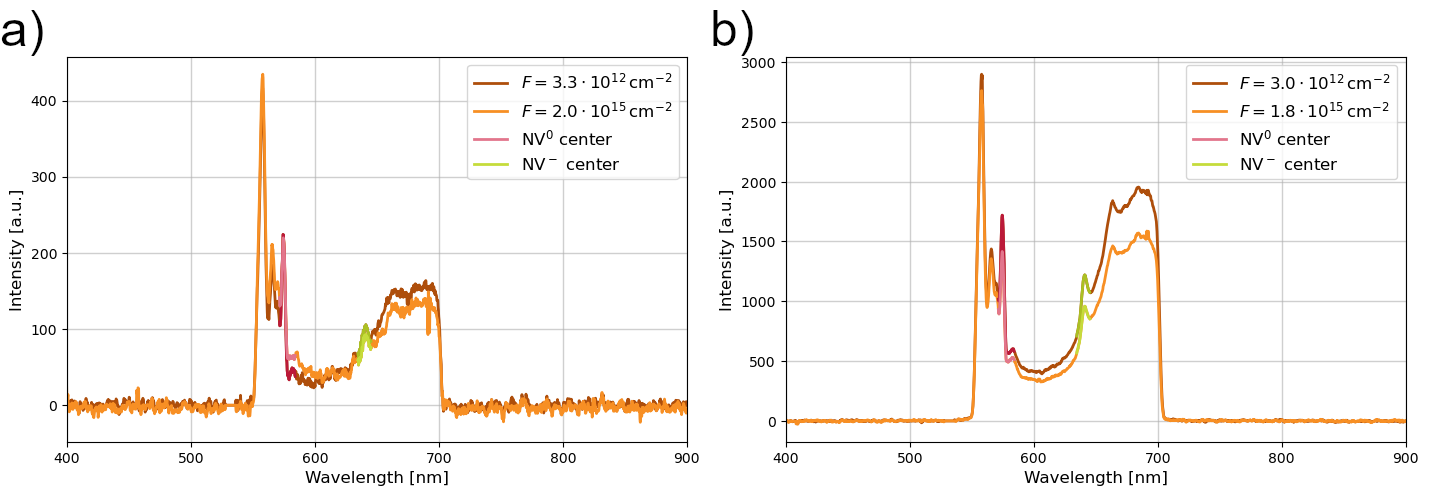}
	\caption{Spectra collected during 2 different IPL runs on a type Ib HPHT diamond:\newline
	\textbf{a)} $I_\mathrm{ion}=50\,\mathrm{pA}$, $P_{laser}=4.2\,\mathrm{mW}$; \textbf{b)} $I_\mathrm{ion}=50\,\mathrm{pA}$, $P_{laser}=32.3\,\mathrm{mW}$.}
	\label{fig6}
	\centering
\end{figure}

In order to resolve NV-related spectral features and to observe the dynamics of NV color centers in type Ib diamonds, the current was set to 50 pA and laser power of 4.2, and 32.3 mW were used. Fig. \ref{fig6} shows spectra from these irradiation runs. Several conclusions can be made from observing the graphs in Fig. \ref{fig6}. First, the laser excitation enables the detection of luminescence signal in the regime of ion currents unavailable to IL measurements, and second the intensity of the collected signal increases for higher laser power. All in all, this demonstrates that the LIBIL system can be used to detect luminescence emission for ion currents below the IL threshold. Additionally, further developments in the light collection system could yield additional increases in detected emission.

\begin{figure}
	\includegraphics[width=\textwidth]{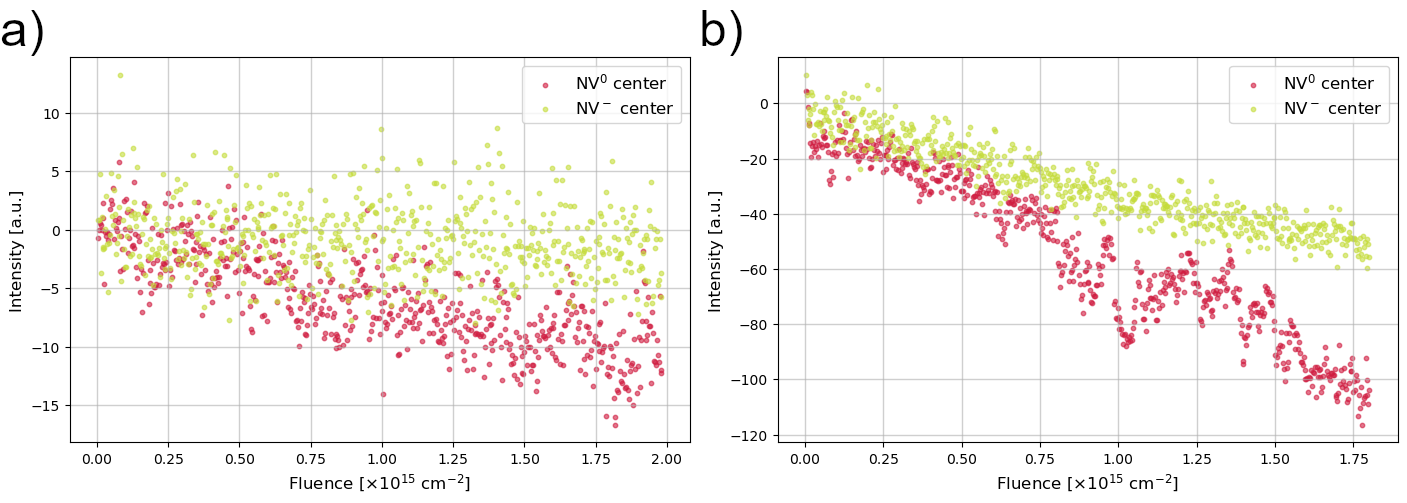}
	\caption{NV$^0$ and NV$^-$ intensities during irradiation for 2 different IPL runs on a type Ib HPHT diamond: \textbf{a)} $I_\mathrm{ion}=50\,\mathrm{pA}$, $P_{laser}=4.2\,\mathrm{mW}$; \textbf{b)} $I_\mathrm{ion}=50\,\mathrm{pA}$, $P_{laser}=32.3\,\mathrm{mW}$.\newline
	The y-axis represents the difference in intensity of collected spectra from the PL spectrum acquired before irradiation.}
	\label{fig7}
	\centering
\end{figure}

The intensity of the NV$^0$ and NV$^-$ peaks was tracked during both irradiation runs and are presented in Fig. \ref{fig7}. The change in observed intensity was different than what is usually described by the Birks-Black model, which could indicate that a more complex model is needed for describing this type of behavior.

\begin{figure}
	\includegraphics[width=\textwidth]{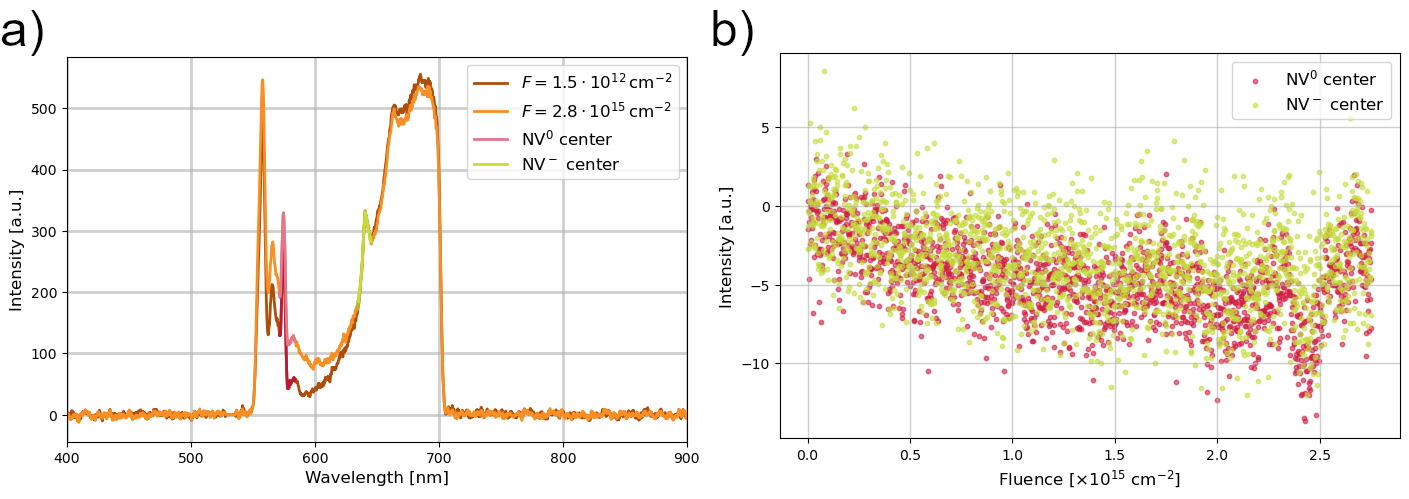}
	\caption{\textbf{a)} Spectra collected on type Ib HPHT diamond for $I_\mathrm{ion}=25\,\mathrm{pA}$ and $P_\mathrm{laser}=32.3\,\mathrm{mW}$; \textbf{b)} Intensity of both NV$^0$ and NV$^-$ peaks during same run. The y-axis represents the difference in intensity of collected spectra from the PL spectrum acquired before irradiation.}
	\label{fig8}
	\centering
\end{figure}

Finally, the ion beam current was further reduced down to 25 pA, while the laser power was kept at 32.3 mW to check if it is possible to further reduce the ion current and still be able to resolve spectral features. The results of this irradiation run are shown if Fig. \ref{fig8}.

\section{Conclusion}

In this work, development and testing of the new Laser and Ion Beam Induced Luminescence (LIBIL) experimental end-station has been presented.

To check if there is any increase in luminescence signal, both ionoluminescence (IL) and iono-photoluminescence (IPL) measurements were performed on the type IIa optical grade diamond with different ion beam currents clearly show that there are spectral features (NV$^-$ zero-phonon line) which are only distinguished with laser excitation present, and the overall emission intensity shows an increase during IPL measurements (see Fig. \ref{fig2}).

Furthermore, the change in emission intensity of the NV$^0$ zero-phonon line with respect to deposited fluence was measured (see Fig. \ref{fig3}). The data was then fitted to the Birks-Black model (Eq. \ref{eq1}), and the parameter $F_{1/2}$ was evaluated (see Fig. \ref{fig4}). The IL results obtained are comparable with the ones in found literature. No significant change in the NV$^-$ intensity was observed. Additionally, it has been shown that the addition of laser excitation doesn't introduce further damage to the material.

Finally, both IL and IPL measurements were performed on a rich type Ib diamond (<200 ppm of nitrogen). The IL results (see Fig. \ref{fig5}) for ion currents of 600 and 400 pA have shown that no NV-related spectral features can be resolved just by IL measurements, and that for 200 pA the setup wasn't able to detect any signal. By adding laser excitation during irradiation, with ion currents well below 200 pA, the setup was still able to detect emission from the irradiated region and resolve spectral features corresponding to both the NV$^0$ and NV$^-$ centers (see Figs. \ref{fig6}, \ref{fig7}, and \ref{fig8}). The intensity of the NV$^0$ and NV$^-$ peaks was tracked during both irradiation runs. The change in observed intensity was different than what is usually described by the Birks-Black model, which could indicate that a more complex model is needed for describing this type of behavior.

Ionoluminescence (IL) is inherently a destructive ion beam analysis (IBA) technique. Therefore, it is difficult to conduct any defect characterization without irreparably damaging the sample. Furthermore, compared to photoluminescence (PL), it is quite inefficient at generating luminescence. Additionally, PL measurements are much less destructive on the sample, making it the preferred method for detection of optically active defects. As shown here, the LIBIL setup  allows for real time monitoring of defect dynamics much like IL, without missing the information acquired from PL measurements (i.e. can resolve spectral features not visible with IL). Currently however, the setup has a major drawback of not being in a confocal configuration. This increases the interaction volume of the laser light, increasing the signal coming from the non-irradiated regions. With further developments (i.e. making a confocal setup), it could be possible to increase the excitation efficiency allowing for detection of more faint luminescence signals. Furthermore, by changing the laser wavelength or output power, the setup could be used for in-situ activation of color centers on a more deterministic scale. The development of such a setup could provide a more efficient and accurate way to characterize color centers during the formation step which could lead to new useful insights into the dynamics of color center creation.

\backmatter


\end{document}